%% file: BullerIsotope.tex
\documentclass[%
 aip,
 jmp,%
 amsmath,amssymb,
 reprint,%
 nofootinbib,%
 floatfix, %
]{revtex4-1}

\usepackage{graphicx}
\usepackage{caption}
\usepackage{subcaption}
\usepackage{dcolumn}
\usepackage{bm}
\usepackage[dvipsnames]{xcolor}
\usepackage{amssymb}
\usepackage{verbatim}
\usepackage{appendix}
\usepackage{listings}
\usepackage{units}
\usepackage{fancyhdr}
\usepackage{csquotes}
\usepackage{hyperref}

\usepackage{etoolbox}
\makeatletter
\patchcmd{\hyper@makecurrent}{%
    \ifx\Hy@param\Hy@chapterstring
        \let\Hy@param\Hy@chapapp
    \fi
}{%
    \iftoggle{inappendix}{
        \@checkappendixparam{chapter}%
        \@checkappendixparam{section}%
        \@checkappendixparam{subsection}%
        \@checkappendixparam{subsubsection}%
        \@checkappendixparam{paragraph}%
        \@checkappendixparam{subparagraph}%
    }{}%
}{}{\errmessage{failed to patch}}

\newcommand*{\@checkappendixparam}[1]{%
    \def\@checkappendixparamtmp{#1}%
    \ifx\Hy@param\@checkappendixparamtmp
        \let\Hy@param\Hy@appendixstring
    \fi
}
\makeatletter

\newtoggle{inappendix}
\togglefalse{inappendix}

\apptocmd{\appendix}{\toggletrue{inappendix}}{}{\errmessage{failed to patch}}
\apptocmd{\subappendices}{\toggletrue{inappendix}}{}{\errmessage{failed to patch}}

\newcommand*\obar[2][0.75]{
    \sbox{\myboxA}{$\m@th#2$}%
    \setbox\myboxB\null
    \ht\myboxB=\ht\myboxA%
    \dp\myboxB=\dp\myboxA%
    \wd\myboxB=#1\wd\myboxA
    \sbox\myboxB{$\m@th\overline{\copy\myboxB}$}
    \setlength\mylenA{\the\wd\myboxA}
    \addtolength\mylenA{-\the\wd\myboxB}%
    \ifdim\wd\myboxB<\wd\myboxA%
       \rlap{\hskip 0.5\mylenA\usebox\myboxB}{\usebox\myboxA}%
    \else
        \hskip -0.5\mylenA\rlap{\usebox\myboxA}{\hskip 0.5\mylenA\usebox\myboxB}%
    \fi}
\makeatother

\definecolor{Gold}{rgb}{1,0.84,0}

\definecolor{C1}{RGB}{51,34,136}
\definecolor{C2}{RGB}{136,204,238}
\definecolor{C3}{RGB}{68,170,153}
\definecolor{C4}{RGB}{17,119,51}
\definecolor{C5}{RGB}{153,153,51}
\definecolor{C6}{RGB}{221,204,119}
\definecolor{C7}{RGB}{102,17,0}
\definecolor{C8}{RGB}{204,102,119}
\definecolor{C9}{RGB}{136,34,85}
\definecolor{C10}{RGB}{170,68,153}

\definecolor{OC1}{RGB}{166,206,227}
\definecolor{OC2}{RGB}{31,120,180}
\definecolor{OC3}{RGB}{178,233,138}
\definecolor{OC4}{RGB}{51,160,44}
\definecolor{OC5}{RGB}{251,154,153}
\definecolor{OC6}{RGB}{227,26,28}
\definecolor{OC7}{RGB}{253,199,111}
\definecolor{OC8}{RGB}{255,127,0}
\definecolor{OC9}{RGB}{202,178,214}
\definecolor{OC10}{RGB}{106,61,154}

\definecolor{OOC1}{RGB}{195,170,60}
\definecolor{OOC2}{RGB}{93,54,134}
\definecolor{OOC3}{RGB}{101,188,103}
\definecolor{OOC4}{RGB}{194,106,187}
\definecolor{OOC5}{RGB}{112,143,57}
\definecolor{OOC6}{RGB}{108,126,215}
\definecolor{OOC7}{RGB}{182,120,55}
\definecolor{OOC8}{RGB}{70,193,154}
\definecolor{OOC9}{RGB}{185,74,115}
\definecolor{OOC10}{RGB}{186,76,65}

\definecolor{OOOC1}{RGB}{94,66,0}
\definecolor{OOOC2}{RGB}{71,103,222}
\definecolor{OOOC3}{RGB}{214,214,70}
\definecolor{OOOC4}{RGB}{85,0,71}
\definecolor{OOOC5}{RGB}{40,213,123}
\definecolor{OOOC6}{RGB}{220,46,88}
\definecolor{OOOC7}{RGB}{5,121,84}
\definecolor{OOOC8}{RGB}{255,169,246}
\definecolor{OOOC9}{RGB}{0,89,41}
\definecolor{OOOC10}{RGB}{250,143,56}

\renewcommand{\vec}[1]{\boldsymbol{#1}}

\newcommand{\e}{\ensuremath{\mathrm{e}}}
\newcommand{\p}{\ensuremath{\partial}}

\newcommand{\DeltaGL}{\ensuremath{\Delta_{\text{GL}}}}

\renewcommand{\d}{\ensuremath{\mathrm{d}}}

\newcommand{\remove}[1]{{}}

\newcommand{\psin}{\ensuremath{\psi_N}}
\newcommand{\nped}{\ensuremath{n_{\rm Ped}}}
\newcommand{\nlc}{\ensuremath{n_{\rm LCFS}}}
\newcommand{\perfect}{\textsc{Perfect}}

\newcommand*{\nouncite}[1]{Ref.~\citenum{#1}}

\newcommand{\dndp}{\protect\scalebox{1}{$\frac{\Delta n}{\Delta \psi_N}$}}

\newcommand{\wstar}{\ensuremath{\delta_n^\ast}}

\newcommand{\na}{\nabla}

\newcommand{\BT}{B_{\rm T}}

\newcommand{\appref}[1]{\hyperref[#1]{Appendix~\ref{#1}}}

\usepackage{morefloats}
\begin{document}

\preprint{AIP/123-QED}

\title{Isotope and density profile effects on pedestal neoclassical transport}
\author{S. Buller} 
\email{bstefan@chalmers.se} 
\affiliation{Department of Physics, Chalmers University of Technology,
  SE-41296 G\"{o}teborg, Sweden}
\author{I. Pusztai} 
\affiliation{Department of Physics, Chalmers University of Technology,
  SE-41296 G\"{o}teborg, Sweden}

\date{\today}

\begin{abstract}
Cross-field neoclassical transport of heat, particles and momentum is studied in sharp density pedestals, with a focus on isotope and profile effects, using a radially global approach. Global effects -- which tend to reduce the peak ion heat flux, and shift it outward -- increase with isotope mass for fixed profiles. The heat flux reduction exhibits a saturation with a favorable isotopic trend. A significant part of the heat flux can be convective even in pure plasmas, unlike in the plasma core, and it is sensitive to how momentum sources are distributed between the various species. In particular, if only ion momentum sources are allowed, in global simulations of pure plasmas the ion particle flux remains close to its local value, while this may not be the case for simulations with isotope mixtures or electron momentum sources. The radial angular momentum transport that is a finite orbit width effect, is found to be strongly correlated with heat sources.
\end{abstract}


\maketitle

\section{Introduction}
\label{sec:intro}

The magnetic fusion community acquired its vast operational experience
and experimental knowledge of stability and confinement predominantly
from deuterium (D) -- and to some extent hydrogen (H) --
discharges. There is much less experience with the reactor relevant
deuterium-tritium mixture (DT), which has only been used in a limited
number of discharges in JET \cite{jacquinot99DT} and TFTR
\cite{strachan97}. Likewise, ITER \cite{ITERPB} will first operate
with He and H \cite{sips15}, and then with D, before starting its DT
operation.  To be confident in our predictions for reactor-scale
devices, such as ITER, we therefore need to understand how various
physical processes are affected by changes in the isotope composition
of the bulk plasma.

In particular, the \emph{isotope scaling} of the energy confinement
has been a long-standing unresolved issue in the field
\cite{dong94,ernst98,bateman99,Tokar04,estradamila05,pusztai11isotope,bustos15,shen16,nakata16,guo17}.
The confinement in the plasma core is observed to be broadly consistent with turbulent transport predictions of gyrokinetic codes, see \nouncite{holland2016} and references therein. Multiscale gyrokinetics -- the theoretical foundation of these codes -- assumes characteristic fluctuation time and length scales that correspond to gyro-Bohm level fluxes.
Na\"{i}ve applications of the gyro-Bohm scaling predict an increased transport with increasing isotope mass. In contrast, the global energy confinement is experimentally observed to improve with increasing isotope mass\cite{bessenrodtWeberplas1993isotopeASDEX,hawryluk98,jacquinotJET99,scott95}.

The exact strength of this favorable isotope scaling varies greatly
between different operational regimes.  As general trends, in Ohmic
and L-mode plasmas the isotope scaling of confinement is weak, while
H-mode D plasmas show a consistently higher confinement time than H
plasmas \cite{schissel89,jacquinotJET99,CMaggi2017JETIsotope}.  Since
the H-mode features an edge transport barrier -- commonly referred to
as the \emph{pedestal} -- such differences between operational regimes
suggest that different mechanisms may be responsible for the isotope
scaling in the core and in the pedestal\cite{cordey99}; the latter
being the focus of this work.

In the pedestal, the plasma parameters can vary significantly over a
thermal orbit width, so that the transport at a given minor radius no
longer can be characterized by local profile values, but depends on
values at nearby radii. Such transport is said to be \emph{radially
  global}, and simple dimensionally motivated or diffusive scaling
estimates, such as gyro-Bohm, may not be appropriate.  The distance
over which the perturbed distribution function is radially coupled
scales with the thermal ion orbit width, which introduces a mass
dependence. In this paper we consider the impact of the radial
coupling on neoclassical heat, particle and momentum fluxes,
particularly in relation to isotope effects and characteristic
pedestal features, such as pedestal height and width.

Specifically, we study how characteristic features of pedestal density
profiles affect the radially global transport fluxes, by varying
pedestal properties of model profiles in simulations with different
isotopes.  A self-consistent modeling of how the profiles themselves
are affected by changes in isotopic composition -- due to e.g.~atomic
physics processes \cite{maggi15,omotaniNF16}, magnetohydrodynamic
stability \cite{snyderNF11} and other processes -- is beyond the scope
of this study.

We solve a global $\delta f$ drift kinetic equation numerically using
the \perfect{}\cite{landreman12,landreman2014} solver.  This tool
represents an intermediate step between the conventionally employed,
core-relevant local $\delta f$ codes \cite{Wong11,BelliNEO}, and the
generally valid, but computationally challenging full-$f$ approach
\cite{ChangLH}. \perfect{} allows density and potential profiles that
vary on ion orbit-width scales, while the linearization around a
Maxwell-Boltzmann distribution requires that the ion temperature
varies weakly over an ion orbit width and that the ion diamagnetic and
$E\times B$ flows nearly cancel in the large gradient region -- a
situation referred to as \emph{electrostatic ion confinement}, which
has been borne out in experimental results
\cite{viezzerASDEXped2015,theiler2017}.

The remainder of this paper is structured as follows: In
\autoref{sec:sim} we discuss the details of our modeling, starting by
explaining relevant aspects of the global $\delta f$ model. We then
discuss the choices made in setting up our model profiles, and finally
provide technical details on the magnetic geometry and
normalization. The presentation and interpretation of simulation
results is done in \autoref{sec:results}. In
\autoref{sec:speciesonflux}, we consider isotope mass effects on the
ion heat flux using a fixed set of baseline profiles, showing
increasing global effects with isotope mass due to an increasing orbit
width. We then consider the impact of changes in the pedestal density
profile in \autoref{sec:varyingprofiles}. We quantify the importance
of global effects on the ion heat flux by two parameters and study
their behavior in terms of profile features and bulk isotopes in
\autoref{sec:peakQ}. We find that the peak ion heat flux is mostly
reduced and it is shifted outwards by global effects, with a larger
possible reduction for heavier isotopes. Finally, we discuss species
and profile effects on particle and angular momentum transport in
\autoref{sec:gammapi}, and correlate momentum transport with heat
sources. Our results are then summarized in \autoref{sec:conc}.

\section{Modeling choices}
\label{sec:sim}

\subsection{Model}
\label{sec:restrict}
We solve the following global $\delta f$ drift kinetic equation
\cite{landreman2014}
\begin{equation}
  \left( v_\| \vec{b} + \vec{v}_{d} \right) \cdot \nabla g
  - C_l[g] = - \vec{v}_m \cdot \nabla f_M +S,
\label{ldkes}
\end{equation}
where $g = f-f_M+(Ze\Phi_1/T) f_M$ is the non-adiabatic perturbed
distribution, $f$ is the gyroaveraged distribution function,
$f_M=[m/(2\pi T)]^{3/2} \eta e^{-W/T}$ is a Maxwellian, $W = mv^2/2 +
Ze\Phi$ is the unperturbed total energy, $m$ and $Ze$ are the charge
and the mass of the species, with $e$ the elementary charge, $\eta= n
e^{Ze\Phi/T}$ is the pseudo-density, $n$ and $T$ are the density and
the temperature of the species, and $\Phi+\Phi_1$ is the total
electrostatic potential with a small perturbed component $\Phi_1$
defined such as to vanish on flux surface average.  The flux surface
average density and pressure of the species are all contained in
$f_M$. In other words, any poloidal variations in plasma parameters
are treated as perturbations. Furthermore, $v_\| =
\vec{v}\cdot\vec{b}$, with $\vec{v}$ the velocity, $\vec{b} \equiv
\vec{B}/B$, with $\vec{B}$ the magnetic field and $B\equiv |\vec{B}|$;
$\vec{v}_{d} = \vec{v}_E + \vec{v}_{m}$ is the unperturbed drift
velocity, $\vec{v}_{m} = v_\|^2 \Omega^{-1} \nabla \times \vec{b}
+v_\perp^2 (2\Omega B^2)^{-1} \vec{B} \times \nabla B$ is the magnetic
drift velocity, $\vec{v}_E = B^{-2} \vec{B} \times \nabla \Phi_0
$ is the E$\times$B drift velocity; $\Omega = ZeB/m$ is the
gyrofrequency, and $\vec{v}_\perp = \vec{v} - v_\| \vec{b}$ is the
velocity perpendicular to $\vec{b}$. $C$ denotes the linearized
Fokker-Planck collision operator.  The gradients are taken holding $W$
and the magnetic moment, $\mu = mv_\perp^2/(2B)$, constant. $S$ is a
source term, which accounts for both real sources (e.g.~ionization,
radiative energy loss) and possible divergence in the fluxes other
than those captured by our solution $g$ (e.g.~a radially varying
turbulent particle flux). We have omitted species indices to
streamline the notation; in cases where this leads to ambiguity, a
lower index $a$ will be used.

Equation~(\ref{ldkes}) is an approximation of the drift-kinetic
equation only when the distribution function is close to a
flux-surface Maxwell-Boltzmann distribution: $f/f_M-1\ll 1$.  This
requires that the temperature $T$ and pseudo-density $\eta$ associated
with the Maxwellian do not vary significantly over a thermal orbit
width, while the density $n$ can vary sharply -- provided that
the electrostatic potential $\Phi$ is such that $\eta$ is slowly
varying. In practice these restrictions are only a concern for the
various ion species; thus, to avoid large deviations from Maxwellian
ion distributions, we consider ion profiles without ion temperature
pedestals. As $T_e$ and $\eta_e$ need only to be slowly varying on
an electron orbit width scale, they are allowed to be comparably sharp to
$n_e\sim n_i$ ($e$ and $i$ subscripts refer to electrons and bulk
ions).  For more details on the self-consistent ordering considered
here the reader is referred to \nouncite{landreman2014,buller2017}.

\subsection{Input profiles}
\label{sec:input}

\begin{figure}
  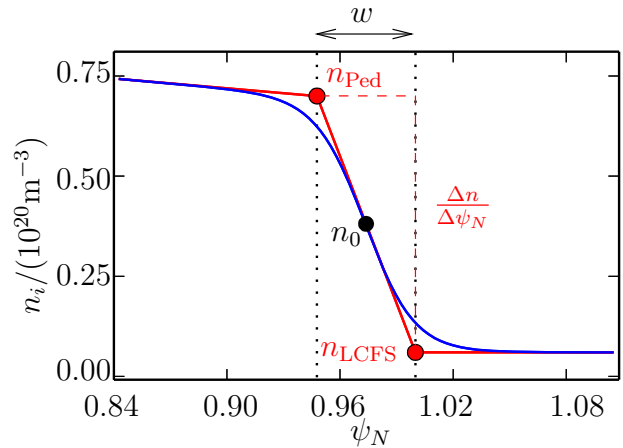
  \caption{\label{fig:input_sketch} The baseline density profile used
    in this work, with profile parameters highlighted. }
\end{figure}

As we have seen in the previous section, even though there are
restrictions on the $T$ and $\eta$ profiles we are free to choose an
arbitrary density profile, which allows us to study the effects of
different density pedestals. For this purpose, we use model density
profiles based on the \emph{mtanh profile}
\cite{Beurskens11,frassinettiNF2017} which is often used to represent
the radial profile dependence of the pedestal.  The mtanh profile
defines three regions: a \emph{core} region, where the density varies
over length scales comparable to the minor radius of the device; a
\emph{pedestal} region, where this variation is comparable to a
typical thermal ion orbit width; and a \emph{SOL} (Scrape-Off Layer)
region with very small gradients (to be discussed further in relation
to boundary conditions at the end of Sec.~\ref{sec:input}); see
\autoref{fig:input_sketch}.

Specifically, a generic mtanh plasma profile $X$ is given by
\begin{equation}
\begin{aligned}
  &X = \frac{1}{2}\left[X_\text{ped} + X_\text{SOL} +
    \frac{w}{2}\left\{\left.\tfrac{\d X}{\d
      \psi_N}\right|_\text{core}-\left.\tfrac{\d X}{\d
      \psi_N}\right|_\text{SOL}\right\}\right] \\ &
  +\frac{\tanh{\left(x\right)}}{2}\left[X_\text{SOL} - X_\text{ped} -
    \frac{w}{2}\left\{\left.\tfrac{\d X}{\d
      \psi_N}\right|_\text{core}+\left.\tfrac{\d X}{\d
      \psi_N}\right|_\text{SOL}\right\}\right] \\ &+
  \frac{\left.\tfrac{\d X}{\d
      \psi_N}\right|_\text{core}(\psi_{N0}-\psi_0)\e^{-x} +
    \left.\tfrac{\d X}{\d
      \psi_N}\right|_\text{SOL}(\psi_N-\psi_{N0})\e^{x}}{\e^{-x} +
    \e^{x}},
 \end{aligned}\label{eq:mtanh}
\end{equation}
where $x=\frac{(\psi_N - \psi_{N0})}{w/2}$ is a radial coordinate,
with $\psin=\psi/\psi_{\rm LCFS}$ being the normalized poloidal flux,
$2\pi\psi$ the poloidal magnetic flux (satisfying $|\nabla \psi|=R
B_{\rm P}$, with $R$ the major radius and $B_{\rm P}$ the poloidal
magnetic field), $\psi_{\rm LCFS}$ is the value of $\psi$ at the last
closed flux surface (LCFS), $\psi_{N0}$ the $\psin$-point in the
middle of the pedestal and $w$ characterizes the pedestal width.
Parameters $X_\text{Ped}$ and $X_\text{LCFS}$ represent the values of
the the asymptotic, linear core and SOL profiles extrapolated to the
pedestal top and the LCFS locations, respectively (marked with red
circles in \autoref{fig:input_sketch}), while $\left.\tfrac{\d X}{\d
  \psi_N}\right|_\text{core}$ and $\left.\tfrac{\d X}{\d
  \psi_N}\right|_\text{SOL}$ are the asymptotic core and SOL
gradients.

In this work, we categorize our density pedestals in terms of four
parameters of common interest: $n_\text{Ped}$, $n_\text{LCFS}$; the
width in $\psi_N$, $w$; and the density gradient in the middle of the
pedestal, $\frac{\Delta n}{\Delta \psi_N}$.  The former three
quantities appear directly in \eqref{eq:mtanh}, while the latter is
defined by the relation
\begin{equation}
n_\text{Ped} + \left(\frac{\Delta n}{\Delta \psi_N}\right) w -
n_\text{LCFS}= 0; 
\label{eq:ped_relations}
\end{equation}
see also \autoref{fig:input_sketch} for a visual definition of the
four pedestal parameters.

We wish to investigate the dependence of transport on these pedestal
parameters.  As the parameters are all related through
\eqref{eq:ped_relations}, we vary two of these parameters at a time,
while keeping the two remaining parameters fixed. This yields 6
different scans.

\begin{figure*}
  \hspace*{-0.2\textwidth}\includegraphics{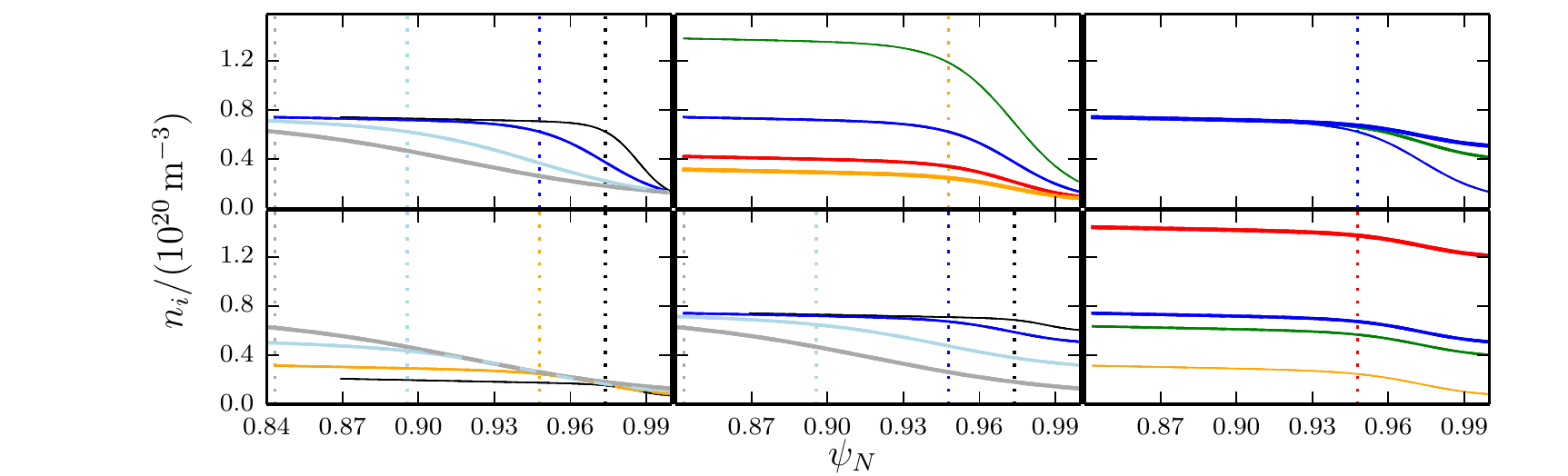}
  \put(-403,135){\large a}
  \put(-272,135){\large b}
  \put(-142,135){\large c}
  \put(-403,75){\large d}
  \put(-272,75){\large e}
  \put(-142,75){\large f}
   \caption{\label{fig:n} Input profiles of $n_i$ in the various
     scans: (a) \dndp{}-$w$, (b) \dndp{}-Ped, (c) \dndp{}-LCFS, (d)
     $w$-Ped, (e) $w$-LCFS, (f) Ped-LCFS. Thinner lines correspond to
     sharper or narrower pedestals in the scan.}
\end{figure*}
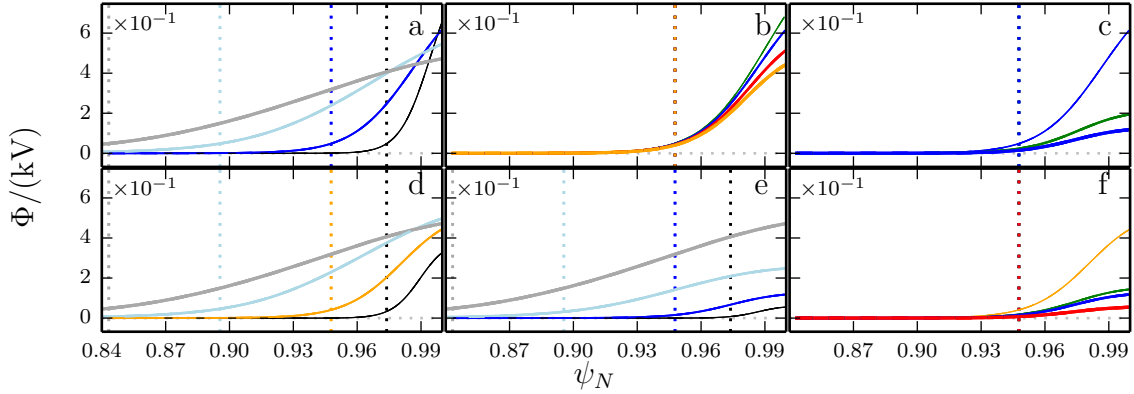
\begin{figure*}
  \hspace*{-0.2\textwidth}\includegraphics{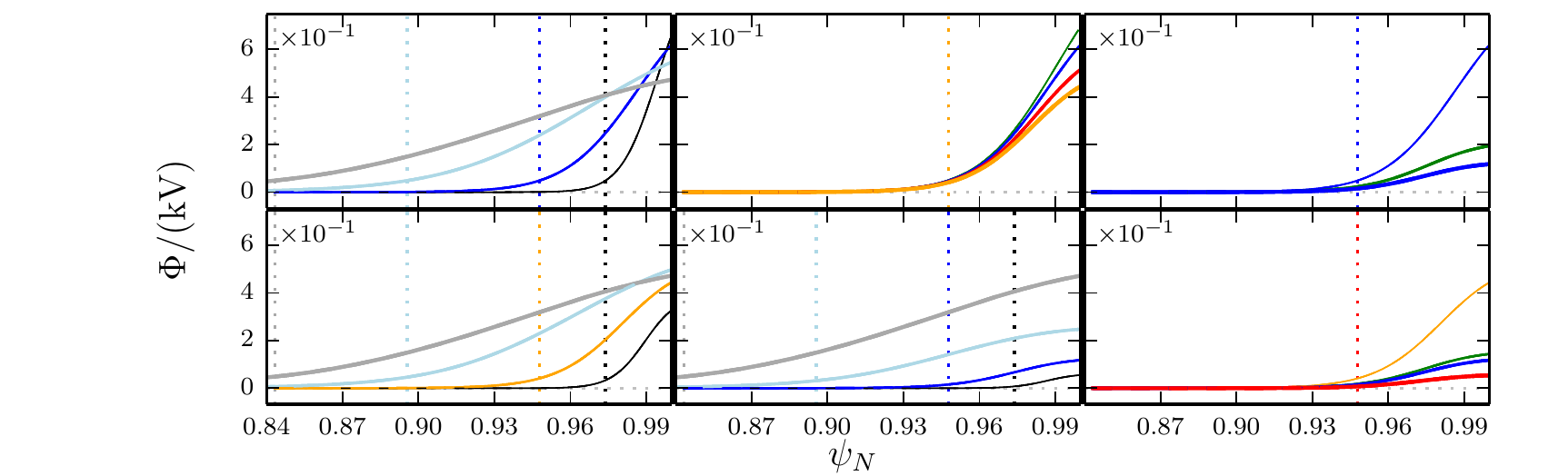}
  \put(-298,135){\large a}
  \put(-167,135){\large b}
  \put(-37,135){\large c}
  \put(-298,75){\large d}
  \put(-167,75){\large e}
  \put(-37,75){\large f}
  \caption{\label{fig:Phi} Electrostatic potentials $\Phi$
    corresponding to the $n_i$ profiles in
    \autoref{fig:n}. Thinner lines correspond to sharper or narrower
    pedestals in the scan.}
\end{figure*}

\begin{table}
   \caption{\label{tab:n} Pedestal parameters corresponding to density
     profiles in \autoref{fig:n}. The scanned parameters are
     written with boldface letters.}
   \begin{tabular}{l|l|l|l}\hline\hline
     $w$ & $\dndp$ & $n_\text{Ped}$ & $n_\text{LCFS}$ \\\hline
\bf 0.0261 & \bf -24.5 & 0.7 & 0.06 \\
\bf 0.0522 & \bf -12.3 & 0.7 & 0.06 \\
\bf 0.104 & \bf -6.13 & 0.7 & 0.06 \\
\bf 0.157 & \bf -4.08 & 0.7 & 0.06 \\
\hline
0.0522 & \bf -24.5 & \bf 1.34 & 0.06 \\
0.0522 & \bf -12.3 & \bf 0.7 & 0.06 \\
0.0522 & \bf -6.13 & \bf 0.38 & 0.06 \\
0.0522 & \bf -4.08 & \bf 0.273 & 0.06 \\
\hline
0.0522 & \bf -12.3 & 0.7 & \bf 0.06 \\
0.0522 & \bf -6.13 & 0.7 & \bf 0.38 \\
0.0522 & \bf -4.08 & 0.7 & \bf 0.487 \\
\hline
\bf 0.0261 & -4.08 & \bf 0.167 & 0.06 \\
\bf 0.0522 & -4.08 & \bf 0.273 & 0.06 \\
\bf 0.104 & -4.08 & \bf 0.487 & 0.06 \\
\bf 0.157 & -4.08 & \bf 0.7 & 0.06 \\
\hline
\bf 0.0261 & -4.08 & 0.7 & \bf 0.593 \\
\bf 0.0522 & -4.08 & 0.7 & \bf 0.487 \\
\bf 0.104 & -4.08 & 0.7 & \bf 0.273 \\
\bf 0.157 & -4.08 & 0.7 & \bf 0.06 \\
\hline
0.0522 & -4.08 & \bf 0.273 & \bf 0.06 \\
0.0522 & -4.08 & \bf 0.593 & \bf 0.38 \\
0.0522 & -4.08 & \bf 0.7 & \bf 0.487 \\
0.0522 & -4.08 & \bf 1.4 & \bf 1.19 \\
\hline\hline
   \end{tabular}
\end{table}

For our baseline density pedestal, we use the pedestal parameters
$\hat{n}_{\text{Ped}} = 0.7$, $w = 0.0522$, $\frac{\Delta
  \hat{n}}{\Delta \psi_N} = -12.3$ and $\hat{n}_{\text{LCFS}} = 0.06$;
these values are comparable to those of the ASDEX Upgrade pedestal shown in
Fig.~1 of \nouncite{wolfrumASDEXped2015}. Here, $\hat{n}$ refers to
the density $n$ given in $\unit[10^{20}]{m^{-3}}$.  Subsequent density
profiles were obtained by scaling two of the pedestal parameters about
the baseline value.  The pedestal parameters for all scans are
displayed in \autoref{tab:n}, with the scanned values written in bold
letters.  From these parameters, we construct mtanh density profiles,
shown in \autoref{fig:n}, where the different sub-figures correspond
to the various scans. These profiles all have the same asymptotic core
density gradient $\left.\frac{\p \hat{n}}{\p
  \psi_N}\right|_\text{core} = -0.40616$ (again, similar to the
corresponding value in Fig.~1 of \nouncite{wolfrumASDEXped2015}), and
$\left.\frac{\p \hat{n}}{\p \psi_N}\right|_\text{SOL}$ is fixed at
zero.

\begin{figure}
  \includegraphics{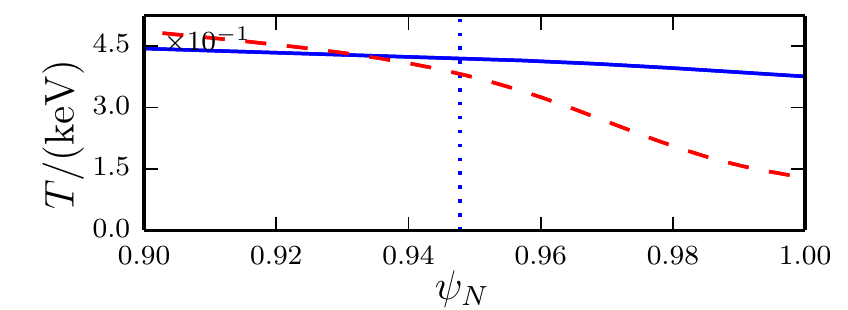}
  \caption{\label{fig:T} The baseline $T_i$ (solid) and $T_e$ (dashed)
    profiles.}
\end{figure}

As we cannot have a pedestal in the ion temperature profile, $T_i$ is
kept fixed at core-like gradients during the scan, as shown in
\autoref{fig:T}.  The profile features a transition in $\na T_i$
centered around the pedestal top ($\psi_N \approx 0.96$ in \autoref{fig:T}), as
this feature is important for the qualitative behavior of both
in-flux-surface and cross-field fluxes\cite{buller2017}, and mimics
the change in gradients at the top of a real pedestal. The asymptotic
gradients are $\left.\frac{\p \hat{T}_i}{\p \psi_N}\right|_\text{core}
= -0.5$ and $\left.\frac{\p \hat{T}_i}{\p \psi_N}\right|_\text{SOL} =
-0.9$, and $\hat{T}_{i,\text{Ped}}=0.42$, where $\hat{T}$ denotes
temperature in units of $\unit[1]{keV}$.

 The baseline electron temperature profile, $T_e$, is an mtanh profile
 with the following parameters: $w=0.0522$, $\frac{\Delta
   \hat{T}_e}{\Delta \psi_N} = -6.12$, $\hat{T}_{e,\text{Ped}}=0.42$,
 and $\hat{T}_{e,\text{LCFS}} = 0.1$. In the scans, we change the
 $T_e$ pedestal width to match that of $n_i$ in each case, while
 $T_{e,\text{Ped}}$ and $T_{e,\text{LCFS}}$ are kept fixed. Similarly,
 the transition in $\nabla T_i$ is moved to match the location of the
 pedestal top, and its width is scaled when the pedestal width is varied. The
 $T_i$ profile is otherwise kept fixed. The baseline $T_e$ and $T_i$ profiles
 are shown in \autoref{fig:T}.

The potential, $\Phi$, is chosen to yield a slowly varying $\eta_i$:
specifically, $\eta_i$ is taken to be an $\rm{mtanh}$ profile
asymptotically equal to the $n_i$ profile in the core, and with core
and pedestal gradients equal to the $n_i$ core gradients -- the
resulting potentials are shown in \autoref{fig:Phi}.  Although there
is some arbitrariness in choosing $\Phi$ this way, it yields an
electric field balancing the ion pedestal pressure gradient --
consistent with typical experimental observations
\cite{viezzerASDEXped2015,ASDEX_Er_from_ion_diamagnetic,theiler2017}.

Although the above profiles have a SOL region, the \perfect{} code
assumes closed field line topology. Rather than a physical SOL, these
regions should be thought of as a numerical buffer zone, such that the
radial boundary conditions we impose on the global drift-kinetic
equation do not affect the results in the pedestal. Accordingly we
do not show results in this region, because they are not physically
meaningful. In this region we set the gradients to be low, so that the
local solution can be used as a boundary condition at both radial
boundaries where particle trajectories enter the domain.

\subsection{Magnetic geometry}
\label{sec:maggeo}

For the magnetic geometry, we use a model Miller equilibrium
\cite{miller98} with the radially constant parameters: $\kappa =
1.58$, $\delta = 0.24$, $s_\delta = 0.845$, $s_\kappa = 0.479$, $\d
R/\d r = -0.14$, $q=3.5$. Here, $\kappa$ is the elongation, $\delta$
is the triangularity, and $s_\kappa$ and $s_\delta$ quantify the
radial variation of these parameters; $R$ is the major radius, $r$ the
minor radius and $q$ the safety factor.  

The radially uniform magnetic geometry, together with the weakly
varying ion temperature profiles, means that the thermal ion
orbit-width is approximately constant throughout the pedestal. Thus,
the radial coupling length of a given species is close to constant in
the simulation domain, providing a simple setting to investigate
global effects.  In reality, magnetic geometry parameters can change
significantly across the pedestal, which would both change the
magnitude and correlation length of the neoclassical transport in the
vicinity of the separatrix. However, our results should be
qualitatively correct in more complicated geometries, as long as the
radial coupling length continues to scale with isotope mass.
  
In scans affecting the pedestal width we keep the LCFS location fixed
and move only the pedestal top location.

\subsection{Normalization of simulation outputs}
\label{sec:norm}

Quantities with a hat are normalized to a reference quantity, $\hat{X}
= X/\bar{X}$.  The reference quantities used in this work are
$\bar{R}=\unit[1.51]{m}$, $\bar{B}=\unit[3.00]{T}$,
$\bar{n}=\unit[10^{20}]{m^{-3}}$, $\bar{T}=e\bar{\Phi}=\unit[1]{keV}$,
$\bar{m}=m^{}_\text{D}$, where $m^{}_\text{D}$ is the mass of deuterium.
From these, we define a reference speed as
$\bar{v}=\sqrt{2\bar{T}/\bar{m}}$, and the dimensionless constant
$\Delta= \bar{m}\bar{v}/(\bar{e}\bar{B}\bar{R})\ll 1$, which is
representative of the normalized gyroradius.  Specific normalizations
are as follows: particle flux, $\hat{\vec{\Gamma}}_a = \int d^3v g_a
\vec{v}_{ma}/(\bar{n}\bar{v})$; radial flux of co-current toroidal
angular momentum (divided by mass), $\hat{\vec{\Pi}}_a = \int d^3v g_a
v_\parallel I \vec{v}_{ma}/(\bar{n}\bar{v}^2\bar{R} B)$, with
$I(\psi)=R\BT$ and  $\BT$ the toroidal magnetic field; heat flux,
$\hat{\vec{Q}}_a =\int d^3v g_a m_a v^2\vec{v}_{ma}/(2
\bar{T}\bar{n}\bar{v})$, that is related to the conductive heat flux
by $\hat{\vec{q}}_a =\hat{\vec{Q}}_a -
(5/2)\hat{T}\hat{\vec{\Gamma}}_a$; and sources $\hat{S}_a =
\bar{v}^2\bar{R} S_a/(\Delta\bar{n}\hat{m}_a^{3/2})$.  In addition, we
define the normalized scalar radial particle flux
\begin{equation}
\hat{\Gamma}_a = \frac{\hat{V}'}{\Delta^{2}\pi\bar{R}\bar{B}} 
  \langle \hat{\vec{\Gamma}}_a \cdot \nabla \psi\rangle, 
\label{normalizedflux}
\end{equation}
and analogously the scalar heat and angular momentum fluxes, where we
introduced the normalized incremental volume $\hat{V}' =
(\bar{B}/\bar{R}) d_\psi V $, with $V(\psi)$ the volume within the
flux surface $\psi$, and the flux surface average is denoted by
$\langle\cdot\rangle$.

\section{Simulation results}
\label{sec:results}

\subsection{Species effects on ion energy flux}
\label{sec:speciesonflux}

In an experiment, changing the bulk ion species causes modifications
to all plasma parameter profiles across the radius. Furthermore,
different quantities can be kept fixed in a species scan (e.g.~total
heating power, average density, etc.), while others change. Here, we
intend to isolate the direct effect of the species from such indirect
effects.  Thus we keep all plasma parameter profiles fixed, as
specified in the previous section (line 2 in \autoref{tab:n}), and
perform both radially local and global simulations with various
hydrogen isotopes: protium (H, which we will simply refer to as
hydrogen), deuterium (D) and tritium (T). Out of these, tritium has
the largest orbit width, and is hence expected to show the largest
differences between the local and global models for a fixed pedestal.

\begin{figure*}
  \hspace*{-0.2\textwidth}\includegraphics{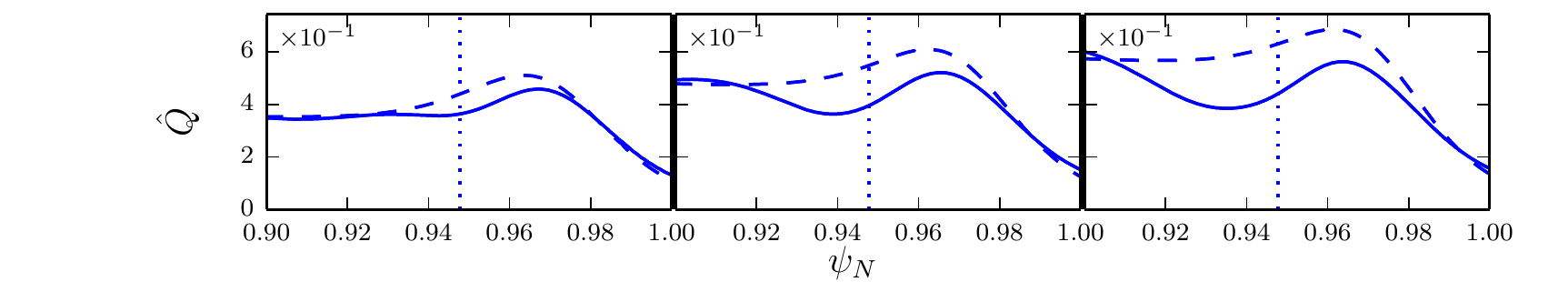}
  \put(-300,67){\large H}
  \put(-170,67){\large D}
  \put(-40,67){\large T}
  \caption{\label{fig:baseQi} Normalized ion heat flux $\hat{Q}_i$ for
    the baseline profile (blue in \autoref{fig:n}) in H, D and T
    plasmas. Solid (dashed) curves are from global (local)
    simulations.}
\end{figure*}

The resulting ion heat fluxes $Q_i$ for the different isotopes are
displayed in the panels of \autoref{fig:baseQi}, where the solid
(dashed) curves correspond to global (local) results, and vertical
lines indicate the nominal location of the pedestal top. As expected,
the deviation between local and global results is the weakest in the H
simulation. This is apparent in two ways: global effects persist for a
shorter distance into the near-pedestal core -- owing to the shorter
radial coupling length that scales with orbit width -- and the
difference at a given radius is typically smaller than for the heavier
isotopes. In particular the global \emph{pedestal peak-value} (PPV),
which is the global maximum value inside the pedestal, is closer to
the local peak value. (The PPV will be used later to characterize
differences in scans. ) In comparison, for the heavier isotopes, the
global PPV is further reduced compared to the local one, and global
effects persist further into the core. 

The deviation between local and global results is not a monotonic
function of the local gradient, rather it appears as a strongly damped
oscillation that dies away as we move further into the core. In this
example it happens to lead to a local minimum in the global $Q_i$ in
the near-pedestal core, for instance, as observed around $\psi_N=0.94$
for $Q_\text{D}$ in \autoref{fig:baseQi}.  Estimating the
coupling-length $r_i$ as the distance from the pedestal top where the
global and local values intersect for the first time, we get
$r_\text{H} = 0.022$, $r_\text{D} = 0.033$, and $r_\text{T} = 0.042$,
which appears to have a mass-scaling $r \propto m^{0.56}$, which
appears consistent with thermal orbit width scaling, $r \propto
m^{1/2}$. Interestingly, the global $Q_i$ inside the pedestal is rather
similar between the three isotopes, even though the local flux
increases due to $\nu_{ii}\rho_i^2 \propto m^{1/2}$ (this random walk
estimate of diffusive transport includes the ion self-collision
frequency $\nu_{ii}$, and the thermal ion Larmor radius $\rho_i$).

\subsection{Varying pedestal profiles} 
\label{sec:varyingprofiles}

\begin{figure*}
  \hspace*{-0.2\textwidth}\includegraphics{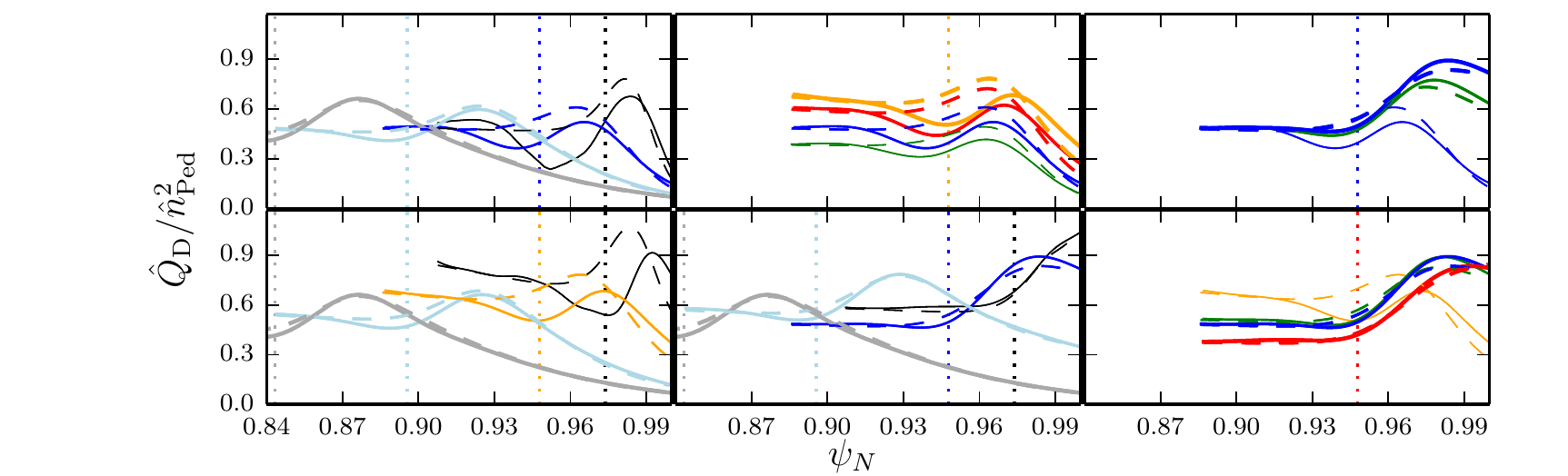}
  \put(-403,135){\large a}
  \put(-272,135){\large b}
  \put(-142,135){\large c}
  \put(-403,75){\large d}
  \put(-272,75){\large e}
  \put(-142,75){\large f}
  \caption{\label{fig:QD} Deuterium heat flux normalized as
    $\hat{Q}_D/\hat{n}_{\rm Ped}^2$ from deuterium simulations using
    the profiles in \autoref{fig:n}. Solid (dashed) curves are from
    global (local) simulations. Thinner lines correspond to
     sharper or narrower pedestals in the scan.}
\end{figure*}
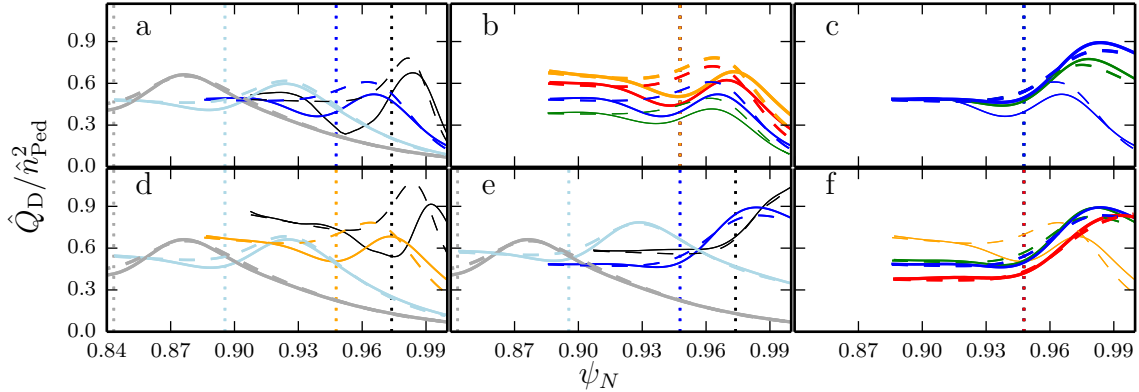

Next, we consider the qualitative changes in the ion heat flux
$\hat{Q}_i$ when the pedestal profiles are modified as in the scans
shown in \autoref{fig:n}.  Here we will keep the isotope fixed.  The
deuterium heat flux $\hat{Q}_{\rm D}$ calculated for the different
input profiles of \autoref{fig:n} is shown in \autoref{fig:QD}, where
again, local simulations are indicated by dashed lines. The
dimensionless heat flux $\hat{Q}$ is divided by $\hat{n}_{\rm Ped}^2$,
to compensate for its expected density scaling in the banana regime,
and so assist the comparison of the results.

In \autoref{fig:QD}a, corresponding to an increasing pedestal width at
fixed pedestal top density, we see how the fluxes become increasingly
more local as the pedestal gets wider and the pedestal gradient
decreases. The result for the narrowest profile (black) yields
significant global effects in the near-pedestal core (with a minimum
at around $\psi_N=0.95$), although the absolute effect on the PPV is
not larger than for the baseline profile result (blue).

\autoref{fig:QD}b corresponds to a pedestal height scan for fixed
pedestal width, producing the same representative pedestal gradients
as those in the \autoref{fig:QD}a scan. One crucial difference is that
while the maximum electric field in the pedestal decreased
significantly as the pedestal width increased, it remains less
affected in this scan (compare the slopes in \autoref{fig:Phi}a and
b), as here $\partial_\psi \ln n$ does not change strongly, except
near the LCFS, since $\nlc$ is kept fixed. As the \autoref{fig:n}b
profiles are more similar to each other in this sense, we see less
pronounced variation in the difference between local and global
fluxes. If anything, the difference is somewhat larger in the cases
with lower pedestal top density, curiously. This may be a consequence
of keeping the ion temperature fixed in this scan: when the global effects due to
the $n$ are reduced, global effects due to $T_i$ may become
increasingly important, and these are perhaps not captured in the
$\nped$ normalization.

In the case shown in \autoref{fig:QD}c the pedestal width and pedestal
top density are kept fixed, while the $\nlc$ is modified leading to a
variation in density gradient. The resulting density gradients are the
same as for the baseline profile and the reduced gradient cases in
\autoref{fig:n}a, with the largest gradient pedestal displaying the largest
global effects.  As the $\nlc$ increases, the maximum of the fluxes --
both local and global -- is shifted outward.

In \autoref{fig:QD}d-e, we study the influence of pedestal width using
profiles with maximum gradients similar to that in the widest pedestal
of \autoref{fig:n}a. In \autoref{fig:QD}d we show simulations with
fixed $\nlc$, which implies that $\nped$ increases with $w$. While the
wide pedestal cases are rather close to local behavior, the thin
pedestal results become surprisingly strongly non-local. This may
partly be due to the large electric field in that case (compare the
slopes in \autoref{fig:Phi}d), and the fact that $\partial_\psi \ln n$
reaches higher values in the small pedestal (as $\partial_\psi \ln n$
is largest in the middle of our model pedestals). Finally, it could
also be related to more abrupt radial changes in the profile that
give rise to strong sources.

However, the latter cause alone is not sufficient to generate large
deviations from the local behavior, as seen from the results keeping
the pedestal top density fixed in \autoref{fig:QD}e. In this scan the
deviations between local and global results are comparable for all
$\nlc$ values, despite the larger (but still small) logarithmic
gradient for the narrower profile.  This similarity could be a result
of the similar electric fields in these cases (consider the slopes in
\autoref{fig:Phi}e).

In the last scan, shown in \autoref{fig:QD}f, we keep the pedestal
width and the gradient fixed, but shift $\nlc$ and $\nped$
simultaneously. The case with the lowest density -- that has the most
realistic pedestal shape -- has the highest $\partial_\psi \ln n$ and
the strongest radial electric field and, as expected, it exhibits the
strongest deviation between the local and global results. As the
density is increased we observe the location of the maximum flux to
shift outward in both the local and the global results. This shift is
caused by the lower relative drop in density across the pedestal in
the higher density cases, noting that the local flux depends on the
local density instead of $\nped$.
 
\subsection[Peak heat flux for various pedestals and isotopes]{Peak heat flux and its location for various pedestals and isotopes}
\label{sec:peakQ}

\begin{figure*}
  \includegraphics[width=1.0\textwidth]{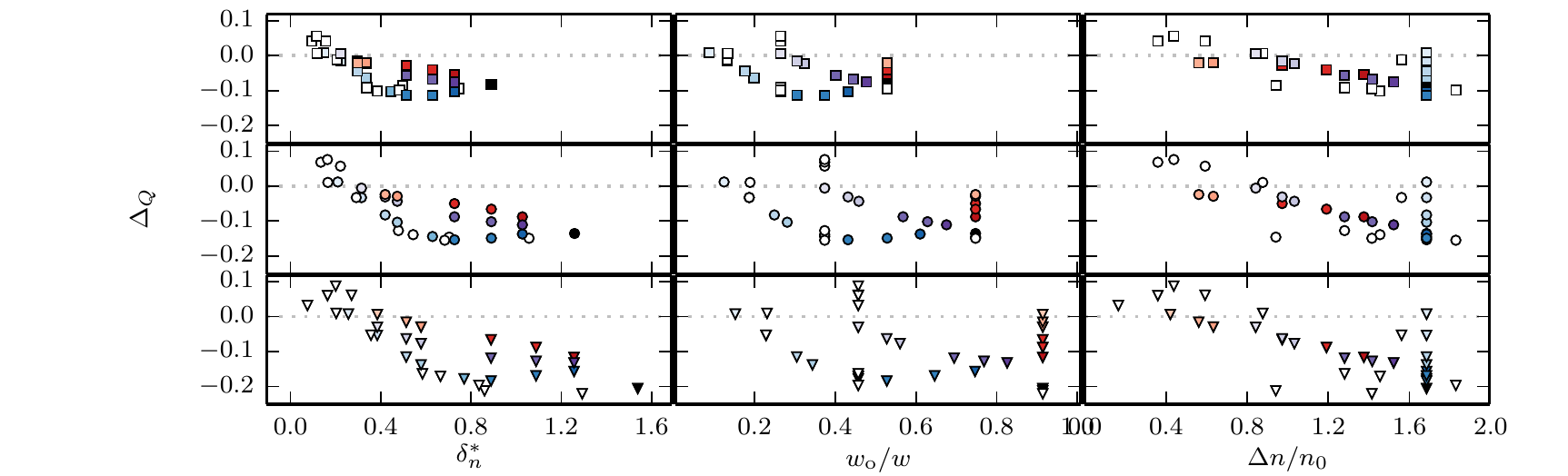}
  \put(-10,95){\large H}
  \put(-10,62){\large D}
  \put(-10,30){\large T}
  \put(-236,105){\large a} 
  \put(-135,105){\large b}
  \put(-34,105){\large c}
  \put(-236,67){\large d}
  \put(-135,67){\large e}
  \put(-34,67){\large f}
  \put(-236,39){\large g}
  \put(-135,39){\large h}
  \put(-34,39){\large i}
  \caption{\label{fig:diffPPV} Relative deviation
    between global and local peak heat flux values in the pedestal. Ion
    species is H (a-c), D (d-f), T (g-i). Representative values of
    (a,d,g) ratio of orbit width and density scale length, (b,e,h)
    ratio of orbit width and pedestal width, (c,f,i) ratio of
    the density drop across the pedestal and the average pedestal density.}
\end{figure*}

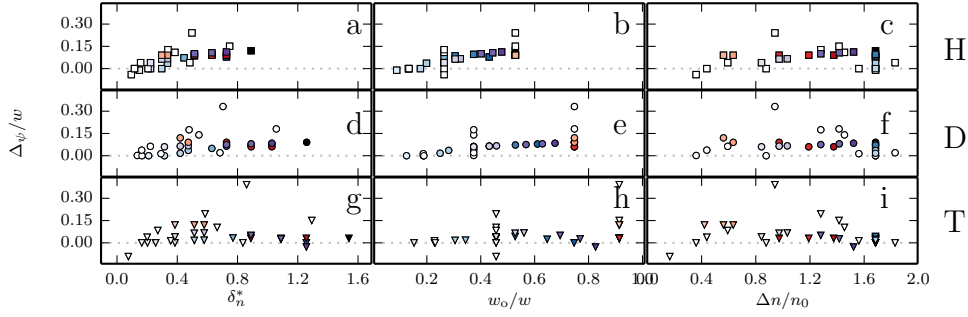
\begin{figure*}
  \includegraphics[width=1.0\textwidth]{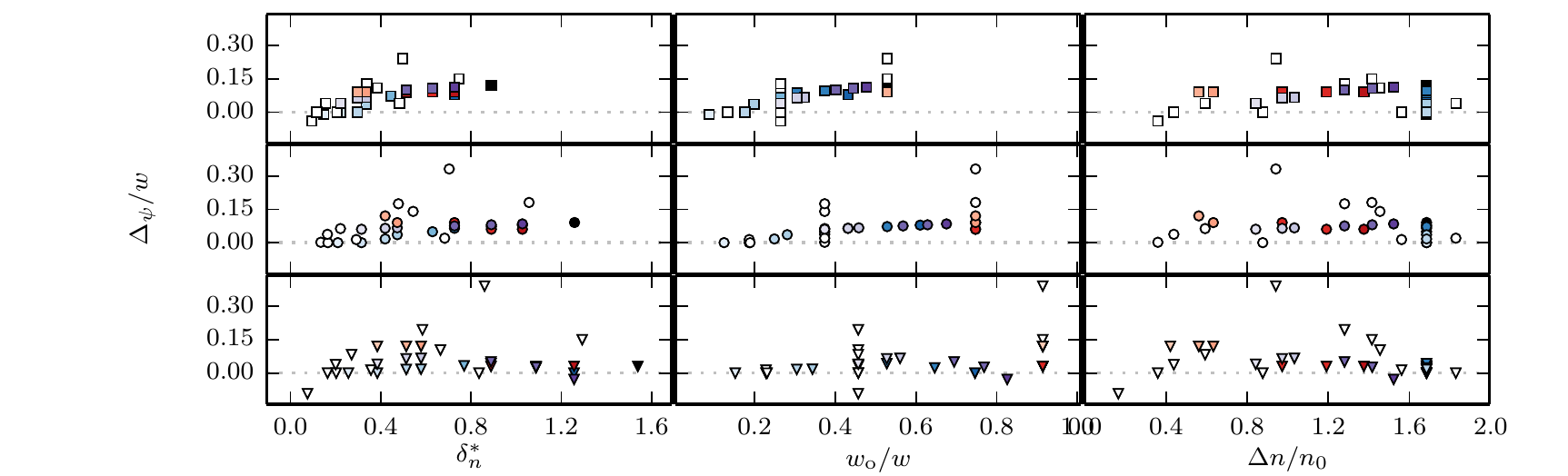}
  \put(-10,95){\large H}
  \put(-10,62){\large D}
  \put(-10,30){\large T}
  \put(-236,105){\large a} 
  \put(-135,105){\large b}
  \put(-34,105){\large c}
  \put(-236,67){\large d}
  \put(-135,67){\large e}
  \put(-34,67){\large f}
  \put(-236,39){\large g}
  \put(-135,39){\large h}
  \put(-34,39){\large i}
  \caption{\label{fig:diffpsiN} Relative deviation between global and local peak
    heat flux locations in the pedestal relative to the pedestal
    width. Ion species is H (a-c), D (d-f), T (g-i). Representative
    values of (a,d,g) ratio of orbit width and density scale length,
    (b,e,h) ratio of orbit width and pedestal width, (c,f,i) ratio of
    the density drop across the pedestal and the average pedestal density.}
\end{figure*}

As we have seen in \autoref{sec:varyingprofiles}, the changes in heat
flux profiles in response to changes in pedestal parameters are rather
complex; in particular the strength of global effects may not follow
intuitively expected trends.  To quantify the profile and isotope
effects identified in the previous sections, we extract the Pedestal
Peak Values (PPVs) of $Q_i$ and the $\psi_N$ location of these
peaks. The significance of the PPV heat flux lies in the fact that it
represents a lower bound on the total heat flux, and that in some
experiments the total ion energy flux is close to neoclassical
predictions \cite{viezzerheatflux}. Since we are mainly interested in
deviations from the local theory, we define $\DeltaGL X = X_\text{G} -
X_\text{L}$: the difference in parameter $X$ between a global and a
corresponding local simulation.  \autoref{fig:diffPPV}, shows the
relative difference between local and global ion heat flux PPVs
$\Delta_Q\equiv (\DeltaGL
\hat{Q}_{i,\text{PPV}})/\hat{Q}_\text{i,\text{PPV},\text{L}}$ for
different ion species (rows of sub-figures) against different
parameters (columns of sub-figures).

Global effects arise when some radial profile length scale
$L_X=-|\nabla \ln X|^{-1}$ becomes comparable with the orbit width
$w_{\rm o}^{(r)}$, that is $\delta_X=w_{\rm
  o}^{(r)}/L_X=\mathcal{O}(1)$. In our case it is the density for
which this happens, thus it is instructive to measure the magnitude of
global effects against $\delta_n$. A representative value of the
maximum $\delta_n$ in the profile is given by $\delta_n^\ast=(w_{\rm
  o}/n_0) \frac{\Delta n}{\Delta \psi_N}$, where $w_{\rm o}=
R\frac{\sqrt{2\epsilon mT}}{Ze \psi_a}$ is a typical trapped orbit
width measured in $\psi_N$ (in contrast with $w_{\rm o}^{(r)}$ that
has the dimension of length), with $\psi_a$ denoting $\psi$ at the
LCFS, and $n_0 = (n_\text{Ped} + n_\text{LCFS})/2$ is the average
pedestal density. Note that the density length scale is typically
comparable to the pedestal width $w$, but these quantities can, in
general, differ significantly, as is the case in our scans. To take an
extreme example, in the scan shown in \autoref{fig:n}f, even though
$w$ and the density gradient are constant, the logarithmic density
gradient, and thus the density scale length, change significantly due
to changes in the density. It is thus also interesting to consider the
effect of the ratio between the orbit width and the pedestal width,
$w_{\rm o}/w$, as another possible indicator of globality.  For
completeness, we also consider the effect of $\Delta n/n_0$. The three
parameters considered are chosen such that they increase for pedestals
where stronger global effects are expected, and can be related to each
other by
\begin{equation}
\delta_n^\ast = \frac{w_{\rm o}}{w} \frac{\Delta n}{n_0}.
\end{equation}

The \wstar{} dependence of the relative variation of the PPV ion heat
fluxes, $\Delta_Q$, is shown in the first column of
\autoref{fig:diffPPV}. Below a certain \wstar{} ($\approx 0.2$, but
increasing with isotope mass) global effects tend to increase $Q_i$
($\Delta_Q>0$), while $\Delta_Q$ remains below $0.1$.  As \wstar{} is
increased, global effects start to lead to a reduction $Q_i$. The
observed $\Delta_Q$ values here occupy a range between $0$ and some
negative envelope. This envelope first increases in magnitude with
\wstar{} then it goes to saturation.  Both the $\wstar{}$ value where
this saturation occurs, and the corresponding maximum relative
reduction in the heat flux, increase with isotope mass. The saturation
is observed at $\wstar\approx 0.3$ ($\wstar\approx 0.6$) with a value
$\Delta_Q\approx -0.12$ ($\Delta_Q\approx-0.22$) for H (T). It is
worth pointing out that the na\"{i}vely expected mass scaling through
the orbit width is already accounted for by defining $\wstar\propto
w_{\rm o}$. That the maximum reduction of $Q_i$ due to global effects
is larger for the heavier isotopes represents a favorable trend with
isotope mass.

Some of the symbols in \autoref{fig:diffPPV} are color coded. The
pedestal with the largest \wstar{} (with the highest logarithmic
gradient) is black. Shades of red correspond to decreasing \wstar{}
from its maximum value by increasing the average pedestal density
while keeping the pedestal gradient and the pedestal width fixed
(similar to \autoref{fig:n}f). In such a scan, $\Delta_Q$ increases
slowly and monotonically as a function of \wstar{}. Shades of blue
correspond to decreasing \wstar{} by making the pedestal wider, while
keeping the pedestal top and the LCFS densities fixed (similarly to
\autoref{fig:n}a). These cases are closer to the lower envelope of the
$\Delta_Q$ range, which exhibits the saturation. Purple shades
correspond to a scan in which both the pedestal width and the average
density are changed, as a superposition of the above mentioned extreme
cases.

It is interesting to examine, whether one of the factors in $\wstar$,
namely $w_\text{o}/w$ or $\Delta n/n_0$, is more strongly correlated
with global effects than the other, or even with $\wstar$. We find
that the correlation of $\Delta_Q$ with $w_\text{o}/w$ is less clear
than it was for \wstar{}, as seen in the second column of
\autoref{fig:diffPPV}. We see that at low values of $w_\text{o}/w$ the
global effects are weaker in magnitude, but in general we find large
scatter in the results for a given value of $w_\text{o}/w$.

Finally we consider the effect of $\Delta n/n_0$, shown
in the last column of \autoref{fig:diffPPV}. As mentioned, at lower
values of this parameter $Q_i$ is increased by the global
effects. When $\Delta n/n_0>0.8$, we mostly see negative
values of $\Delta_Q$ with a large scatter and no clear trends.

As another measure of global effects besides $\Delta_Q$, we may also
consider the difference in the $\psi_N$ locations of the peaks between
global and local heat flux results, which we denote by
$\Delta_\psi$. Unless there is a substantial radiative energy loss in
the pedestal, with sharp radial variation, the location of the maximum
neoclassical heat flux should approximately coincide with the region
of strongest reduction in the ion scale turbulent fluctuation levels,
which in turn, may be measurable with certain fluctuation
diagnostics. Positive values of $\Delta_\psi$ indicate that the global
neoclassical heat flux peaks at a larger radius than the local
flux. To be able to sensibly compare effects in pedestals with
different width $w$ we show $\Delta_\psi/w \in [-1,1]$ in
\autoref{fig:diffpsiN}.

We find that the location of the PPV heat flux is shifted outward in
most cases, as seen in the first column of
\autoref{fig:diffpsiN}. However, there are a few cases when it is
shifted inwards.  With increasing \wstar{} the positive range of
$\Delta_\psi/w$ values increases up to the \wstar{} where the
saturation of $\Delta_Q$ occurred, then the range shrinks again. As
the \wstar{} point where the trend in the $\Delta_\psi/w$ changes
increases with isotope mass, the maximum possible $\Delta_\psi/w$
value is also significantly larger for heavier isotopes.

The second column of \autoref{fig:diffpsiN} shows the $w_0/w$
dependence of $\Delta_\psi/w$. Although in most of the cases we see
only a weak positive shift, there are significant shifts of the heat
flux peak location towards the separatrix; and the range of
$\Delta_\psi/w$ keeps increasing with $w_0/w$. This approximate
proportionality of the envelope of the $\Delta_\psi/w$ data to
$w_0/w$ is related to the appearance of the $1/w$ factor in both
quantities. Reducing the pedestal width (i.e.~increasing $w_0/w$) can
sharpen the pedestal, which would intuitively make non-local effects
stronger, while the observed increase in the maximum relative shift
$\Delta_\psi/w$ is consistent with the trivial effect from the $1/w$
factor.

The effect of $\Delta n/n_0$ on $\Delta_\psi/w$ is shown in the third
column of \autoref{fig:diffpsiN}. We see a few inward shifts of the
peak $Q_i$ location, but mostly find outward shifts. We observe the
strongest outward shift for all species somewhat above $\Delta
n/n_0=0.94$, and at higher values of $\Delta n/n_0$ we see a
decreasing trend in the maximum outward shifts. Interestingly the rate
of this reduction is faster in heavier isotopes. The largest outward
shift is only $\approx 2$ times higher for H than the shift in
the largest \wstar simulation at $\Delta n/n_0\approx 1.68$ (black symbols),
in contrast to the factor of $\approx 11.9$ difference between the corresponding
pair of points for T.  This may relate to the curious fact that when
the pedestal width is decreased for fixed $\Delta n/n_0\approx 1.68$
(symbols with blue shades) the relative outward shift of the $Q_i$
peak is increasing for sharper pedestals for H, while for T the
$\Delta_\psi/w$ points in this scan almost overlap, due to the lower
envelope.

\begin{figure}
  \includegraphics{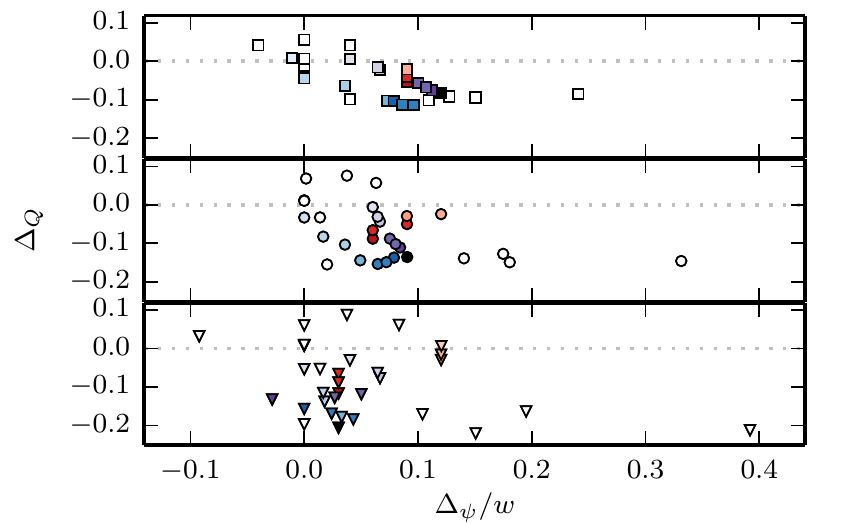}
  \put(-30,122){\large H}
  \put(-30,80){\large D}
  \put(-30,40){\large T}
  \caption{\label{fig:diffQPPV_vs_psiN} $\Delta_Q$ plotted against
    $\Delta_\psi/w$ for different isotopes.}
\end{figure}

Next we consider possible correlations between $\Delta_Q$ and
$\Delta_\psi/w$; these quantities are plotted against each other in
\autoref{fig:diffQPPV_vs_psiN}. Although we observe that $\Delta_Q$
tend to be negative for the highest values of $\Delta_\psi/w$, and
positive for the highest negative values of $\Delta_\psi/w$, we do not
see a clear correlation between these quantities. It is interesting
though, that the three color coded scans show different behaviors for
the various species. In particular the highest \wstar{} pedestal
(black symbols) becomes close to the observed maximum $\Delta_Q$ value
when going from H to T, while $\Delta_\psi/w$ is significantly
reduced.

\subsection{Particle and momentum fluxes}
\label{sec:gammapi}

Unlike in the plasma core, due to  the presence of strong gradients,
neoclassical particle transport can be non-negligible in the pedestal,
in the sense that $T\Gamma_i \sim Q_i$, even if impurities are only
present in trace quantities.

\begin{figure*}
  \hspace*{-0.2\textwidth}\includegraphics{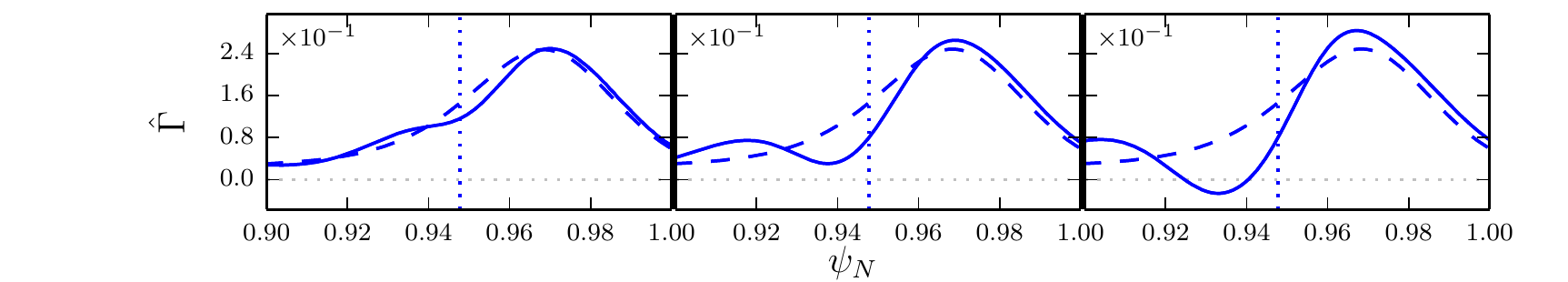}
  \put(-300,67){\large H}
  \put(-170,67){\large D}
  \put(-40,67){\large T}
  \caption{\label{fig:baseGamma} Normalized particle flux
    $\hat{\Gamma}$ for the baseline profile (blue in \autoref{fig:n})
    in H, D and T plasmas. Solid (dashed) curves are from global
    (local) simulations.}
\end{figure*}

In previous studies
\cite{landreman2014,pusztai2016perfect,buller2017}, the global drift
kinetic equation (\ref{ldkes}) has usually been solved together with
the constraints $\langle \int d^3v g\rangle=\langle \int d^3v \,v^2
g \rangle=0$, to obtain the radial dependence of two components of $S$,
corresponding to particle and energy sources. However, $S$ did not
contribute to the angular momentum balance.  This method, in general,
does not lead to ambipolar fluxes, $\sum_a Z_a e\, \Gamma_a=0$. In the
current work we allow for an additional component in $S$,
corresponding to an angular momentum source, and this degree of
freedom is used to enforce ambipolarity, as explained in \appref{app:sources} and in Appendix C of Ref.~\citenum{buller2017}.
Accordingly, the particle fluxes shown in
\autoref{fig:baseGamma} are the same for ions and electrons. Note that
the peak particle fluxes reach values of $\hat{\Gamma}\approx 0.31$,
that is indeed comparable to $\hat{Q}$ in \autoref{fig:baseQi}. Thus,
a significant fraction of the energy flux is convective (while it is
often assumed to be dominantly conductive
\cite{pusztai10,kagan10PPCF,catto13}).

Like the energy flux, the deviation between local and global particle
fluxes increases with isotope mass. While the global effects are quite
weak for H, we find a non-negligible increase in the peak value of
$\hat{\Gamma}$, and an even more pronounced reduction inside the
pedestal top for T. Thus, a significant part of the global reduction
of $\hat{Q}$ in the near-pedestal core is due to a reduced convective
heat flux, while the reduction of $\hat{Q}$ in the pedestal
corresponds to a slight increase in convective and a stronger decrease
in conductive heat flux. The tritium particle flux even changes sign
at $\psi_N\approx 0.94$. Note that the global modifications to the
particle flux are similarly strong for electrons, since the fluxes are
ambipolar. This strong non-local behavior for electrons depends on
whether and how ambipolarity is enforced, i.e.\ it is not a result of
direct orbit width effects, which are negligible for electrons. The
question of ambipolarity and momentum sources in relation to the
particle transport are discussed in \appref{app:sources}. The specific
choices made here are not expected to have major impact on the results
of the previous section, as the peak value of the particle flux is
less sensitive than the that of the ion heat flux with respect to the
ion mass; that is, the isotope dependence of the heat flux is
dominated by the conductive component.

\begin{figure}
  \includegraphics{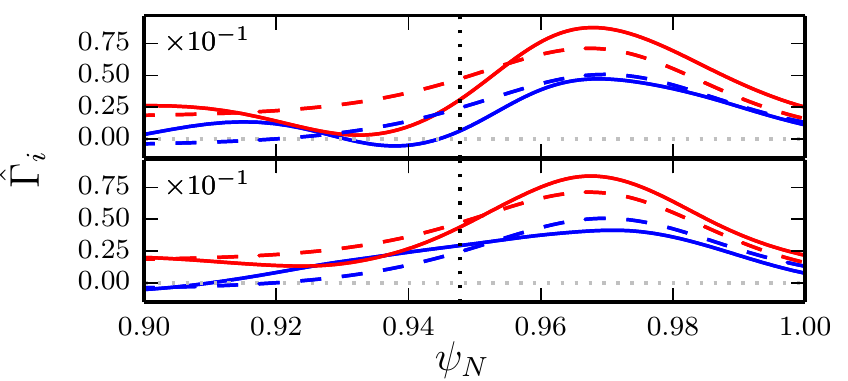}
  \put(-30,92){\large a}
  \put(-30,47){\large b}
  \caption{\label{fig:mixGamma} Normalized particle fluxes
    $\hat{\Gamma}$ for the baseline profile with 1:1 D-T mixture. Blue
    (red) are deuterium (tritium) fluxes. (a): momentum sources
    proportional to particle mass; (b): ambipolarity enforced without
    electron momentum sources, ion momentum sources proportional to
    their mass.  Solid (dashed) curves are from global (local)
    simulations.}
\end{figure}

Nevertheless, different isotopes having different particle fluxes has
potential implications for mixtures in the form of isotope
separation. For instance the ion concentrations in a D-T fusion plasma
can shift away from the ideal 1:1 ratio. That different isotopes in a
mixture are transported differently has long been known, see
e.g.\ \nouncite{connor1973} for analytic theory relevant to the local banana-regime.

To investigate how this is modified by global effects, we performed
simulations of H-D and D-T 1:1 mixtures, with the ion densities taken
as half the baseline electron density. \autoref{fig:mixGamma} depicts the resulting particle fluxes. As the global fluxes depend on how ambipolarity is restored, we performed simulations with electron sources, similar to the pure simulations (\autoref{fig:mixGamma}a), and without electron sources (\autoref{fig:mixGamma}b).
We first consider the local particle fluxes (dashed lines, which are
the same in both \autoref{fig:mixGamma}a and b). In the core the
electron flux is small, thus the ion components are transported in
opposite directions; the D (T) flows are inward (outward). As the
electron profile gradients increase towards the pedestal and
$\Gamma_e$ becomes non-negligible (and outward) both D and T are
transported outward but on different rates; the T flux being
larger. Finally, as the ion collisionality decreases towards the edge
(due to the strong density drop and flat ion temperature profile), the D
and T fluxes get closer to each other.

Sufficiently far from the pedestal the global simulation results
approach the local ones, while closer to the pedestal ($\psi_N \approx 0.92$) the deuterium and tritium fluxes approach each other, and then diverge from each other. Inside most of the pedestal the T flux is even larger than that in the local simulation, while the D flux is somewhat lower, increasing the disparity between these fluxes. These features occur independently of whether electron momentum sources are allowed or not.
In the simulations with electron momentum sources the
individual ion fluxes for the two isotopes resemble the corresponding
single species results; compare to the last two panels of
\autoref{fig:baseGamma}). In the simulation with no electron momentum
sources, they should add up to produce the approximately local
electron result, therefore the deviations from their local results has
to mirror each other. This has the most significant effect in the near
pedestal core, especially for the D flux: while it changes sign around
$\psi_N \approx 0.94$ with electron momentum source, it remains
positive without it. Regardless of these details, as the isotopes are
transported differently in both local and global simulations with a
larger outward T flux, they are prone to develop different density
profiles in experiments, and their fueling may need to be adjusted to
optimize the isotope ratio in the deep core.

\begin{figure*}
  \hspace*{-0.2\textwidth}\includegraphics{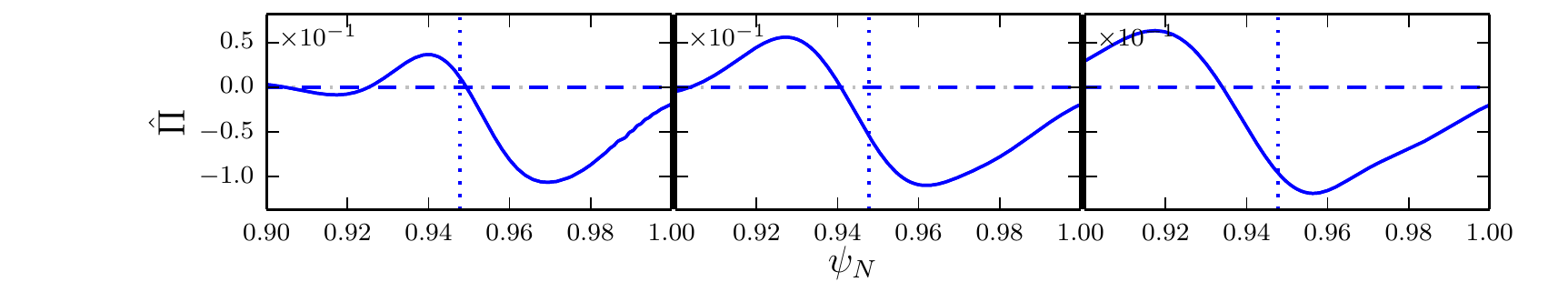}
  \put(-300,67){\large H}
  \put(-170,67){\large D}
  \put(-40,67){\large T}
  \caption{\label{fig:basePi} Normalized total angular momentum flux
    $\hat{\Pi}$ for the baseline profile (blue in \autoref{fig:n}) in
    H, D and T plasmas. Solid (dashed) curves are from global (local)
    simulations.}
\end{figure*}

As for the angular momentum transport, it cannot be sensibly evaluated
in the lowest order local theory as it is a higher order
effect. Indeed, in our local simulations the angular momentum
transport evaluates to zero. However, as it was pointed out in
Ref.~\citenum{eps2016_4page}, momentum transport in the pedestal due
to finite orbit width effects is not only non-negligible, but it
translates to experimentally relevant Prandtl number estimates.

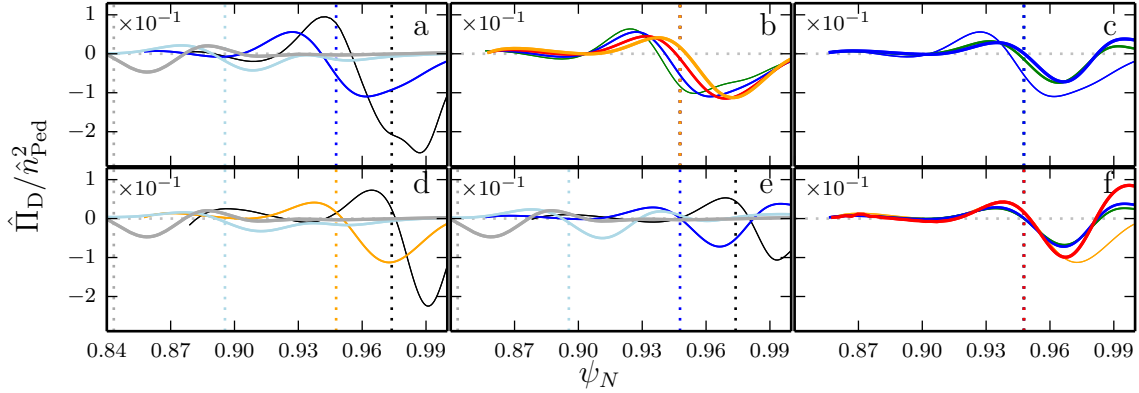
\begin{figure*}
  \hspace*{-0.2\textwidth}\includegraphics{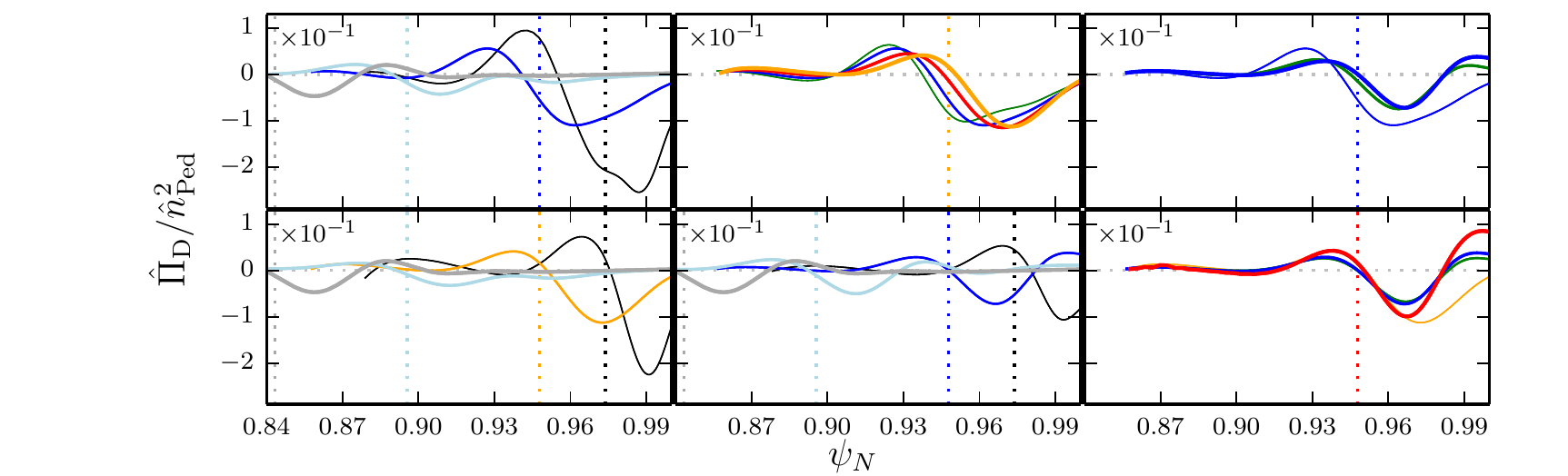}
  \put(-298,135){\large a}
  \put(-167,135){\large b}
  \put(-37,135){\large c}
  \put(-298,75){\large d}
  \put(-167,75){\large e}
  \put(-37,75){\large f}
  \caption{\label{fig:PiD} Angular momentum flux normalized as
    $\hat{\Pi}_D/\hat{n}_{\rm Ped}^2$ from deuterium simulations using
    the profiles in \autoref{fig:n}. Solid (dashed) curves are from
    global (local) simulations. Thinner lines correspond to
     sharper or narrower pedestals in the scan.}
\end{figure*}

First we consider the isotope scaling of momentum transport for our
baseline profiles, shown in \autoref{fig:basePi}. As expected, the
local values are identically zero, while we find an inward flux of
co-current (i.e.~ion-diamagnetic direction) angular momentum inside
the pedestal, and a somewhat weaker outward flux in the near-pedestal
core. In the simulations the radial variation of these momentum fluxes
is balanced by momentum sources. (Note that qualitatively similar
momentum fluxes were reported \cite{buller2017} in simulations without
momentum sources, where a torque from finite radial currents played
the role of momentum sources). Observe that, according to our
definition of $\hat{\Pi}$ given above Eq.~(\ref{normalizedflux}), the
angular momentum transport is $\propto m \hat{\Pi}$.  The toroidal ion
flow can be viewed as a drive for the momentum flux, and its $E\times
B$ and diamagnetic components are mass independent. Nevertheless, it
is somewhat unexpected that the largest negative value of $\hat{\Pi}$
in the pedestal does not increase significantly in magnitude with
increasing isotope mass, since the momentum flux we observe is a
finite orbit width effect.  However, the largest positive value of
$\hat{\Pi}$, inward of the pedestal top, does increase in both
magnitude and its extent towards the core.

Finally, the effect of varying the pedestal parameters on angular
momentum transport is considered; $\hat{\Pi}/\hat{n}_{\rm Ped}^2$ is
shown in \autoref{fig:PiD} for the various density profile scans of
\autoref{fig:n} in a D plasma. The strongest momentum transport is
observed for the highest \wstar{} case, corresponding to the narrowest
pedestal in \autoref{fig:n}a. In this scan, where the position of the
pedestal top is shifted inward, we observe a rather strong reduction
in the largest negative value of $\Pi$ (in the following discussion we
will simply refer to it as the peak $\Pi$) as the gradient
decreases. The location of the peak is shifted outward so that it
stays in the vicinity of the pedestal top; in the shallower pedestals
$\Pi$ is small in most of the pedestal, except close to the pedestal
top. We note that for the model $T_i$ profile used here, the abrupt
increase of $dT_i/d\psi$ close to the pedestal top usually leads to
more pronounced global effects in that region, correlated with a
peaking of sources.  We observe a localization of momentum sources
around the pedestal top in all scans when the pedestal width is
changing; see also \autoref{fig:PiD}d and e.

\begin{figure}
  \includegraphics{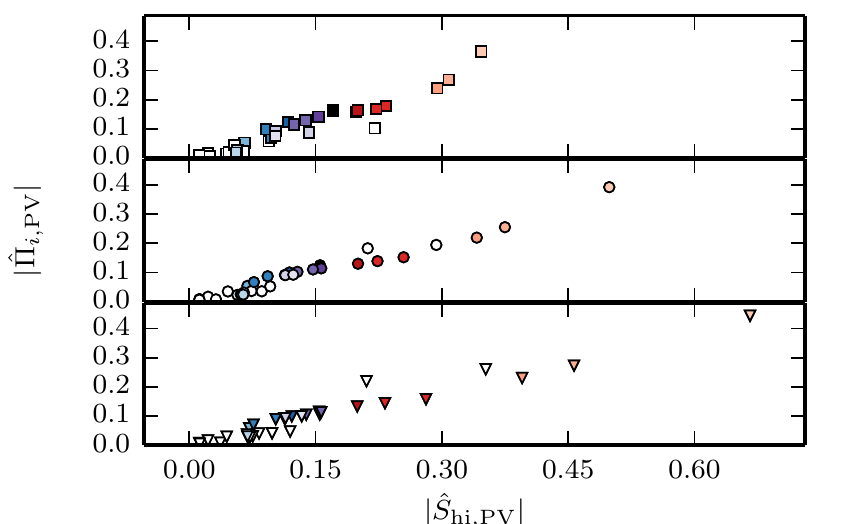}
\put(-30,122){\large H}
  \put(-30,80){\large D}
  \put(-30,40){\large T}
  \caption{\label{fig:Pi_vs_Sh} Pedestal peak values of the momentum
    flux $\hat{\Pi}$ and the heat source $S_h$.}
\end{figure}

The $n^2$ scaling of the local heat fluxes in the banana regime was a
useful guide to choose a convenient normalizing factor,
$1/\hat{n}_{\rm Ped}^2$, in \autoref{fig:QD}. For momentum fluxes such
simple guidance is not available, as the lowest order local momentum
fluxes are zero. However, whilst heat sources (in the form of a radial
variation of turbulent transport fluxes and actual heat sources) may
be expected to be in proportion to the heat fluxes, heat sources will
also represent an important contribution to finite orbit width
momentum fluxes, as shown in the higher order theory of
Ref.~\citenum{ParraTheory2015}). Therefore it is reasonable to assume
an $n^2$ scaling for $\Pi$ as well; indeed, with the $1/\hat{n}_{\rm
  Ped}^2$ normalization we observe comparable peak values of the
momentum transport over the wide range of $\hat{n}_{\rm Ped}$ values
in the scans of \autoref{fig:n}b and f, as seen in the corresponding
\autoref{fig:PiD}b and f. In particular, apart from an inward shift of
the peak $\Pi$ location as $\hat{n}_{\rm Ped}$ increases, we find very
similar peak values of $\hat{\Pi}/\hat{n}_{\rm Ped}^2$ in
\autoref{fig:PiD}b.

To more clearly show that the momentum fluxes and the heat sources are
strongly correlated we plot the largest observed momentum fluxes
against the largest sources (considering only the magnitude of these
quantities) for all our simulations in \autoref{fig:Pi_vs_Sh}; here,
we plot the \emph{peak value} (PV), which differs from the PPV in that
the peak is allowed to be outside the pedestal.  We find an
approximate proportionality with a slope that is almost mass
independent [linear fits give coefficients of $0.86\pm 0.06$ (H),
  $0.68\pm 0.04$ (D), and $0.67\pm 0.04$ (T)]. We have normalized
$\hat{\Pi}$ and $\hat{S}$ such that this mass independence translates
to $\Pi \propto m \int d^3 v \, mv^2 S$, where $\Pi$ is the radial
transport of angular momentum and $S$ is the source appearing in the
kinetic equation (\ref{ldkes}), that conforms qualitatively with the
source contribution in Eq.~(104) of
Ref.~\citenum{ParraTheory2015}. This is potentially important, because
sources in the pedestal are much larger than in the core, as the
radial variation of the fluxes in the various transport channels
happens on a short radial scale.

\section{Summary}
\label{sec:conc}

We have studied cross field fluxes of heat, particles and momentum in
sharp density pedestals. We focused on differences between radially
\emph{local} and \emph{global} simulation results. The deviations
between local and global results depend on isotopic mass, and on
features of the plasma parameter profiles. This study considers
isotope effects for fixed pedestal profiles, and the impact of
changing the shape of the density pedestal. We parametrize the
pedestal by four parameters -- the densities at the pedestal top and
at the last closed flux surface, the representative pedestal gradient,
and the pedestal width -- three of which are independent.

For both particle and heat fluxes we see a clear increase in global
effects with increasing isotope mass, owing to an increasing thermal
ion orbit width. Not only does the deviation between local and global
results increase inside the pedestal, but so does the distance over
which global effects penetrate into the core. The global particle flux
is found to be particularly sensitive to isotope effects: when electron momentum sources are allowed the tritium
particle flux is strongly reduced in the near-pedestal core region,
even changing sign, due to global effects. When only the ions are allowed to have momentum sources, the global particle flux remain close to the local one. However in isotope mixtures and impure plasmas the various ion components may exhibit deviations from local results, as long as ambipolarity is satisfied.

Even though the angular momentum transport is purely a finite orbit
width effect (it vanishes in the lowest order local theory), it does
not exhibit a strong isotopic dependence, apart from its trivial
proportionality with mass (in the sense that, if momentum diffusivity
could be sensibly defined it would approximately be mass independent).

In a scan where the pedestal width and the LCFS density are fixed and
the pedestal top density is increased, we see only weak changes in
globality.  Perhaps more surprisingly, we find that in a scan where
the pedestal top density and location is changed keeping the
representative gradient in the pedestal fixed, thin pedestals can be
much more global than wider ones. Although we only considered steady
state profiles, this scan may be relevant in the growth phase of
ballooning mode limited pedestals, when the large gradient region
gradually expands into the core, while the pedestal pressure gradient
is approximately constant. These observations indicate the role
of the density length scale in affecting globality.

Since experimental ion heat fluxes have been found to be consistent
with being purely neoclassical in some cases, the peak ion heat flux
and its location within the pedestal are of interest. The relative
changes in these two quantities due to global effects are used here to
quantify globality.  We studied their dependence on three pedestal
shape parameters: the ratio of the ion orbit width and the density
length scale, \wstar{}, the ratio of the orbit width and the pedestal
width, $w_{\rm o}/w$, and the relative density drop in the pedestal
$\Delta n/n_0$. We found that, the peak ion heat flux is mostly
reduced and it is shifted outward by global effects (these are modest
changes but they might be measurable). The clearest correlation was found
with \wstar{}. Even though \wstar{} accounts for the mass dependence
through the ion orbit width, we find additional differences between
the various isotopes. The range of possible reductions in the heat
flux peak saturates at some value of \wstar{}. Importantly, this value
and the corresponding maximum reduction of the heat flux increases
with mass, representing a favorable isotope scaling trend. Finally, we
find that the flux of radial angular momentum is strongly correlated
with the heat sources that appear in the simulation to sustain the
steady state pedestal profiles.

\acknowledgments We are grateful to S.L. Newton and J.T. Omotani for
valuable comments on our manuscript, and to M. Landreman for providing
the \perfect{} code. The research was supported by the International
Career Grant of Vetenskapsr{\aa}det (Dnr.~330-2014-6313) and Marie
Sklodowska Curie Actions, Cofund, Project INCA 600398.  The
simulations used SNIC computational resources at Kabnekaise (Dnr:
2017/3-29) and Hebbe (Dnr: 2017/1-95).

\appendix

\section{\label{app:sources} Momentum sources and ambipolarity}
In the local theory $\delta f = f-f_M\ll f_M$ and $\nabla \ln \delta f
\sim \nabla \ln f_M$, thus $\nabla \delta f \ll \nabla f_M$, while in
the global theory adopted here $\delta f$ can develop sharp features
radially, so that $\nabla \delta f \sim \nabla f_M$. Accordingly,
$\nabla \delta f$ needs to be retained one order lower than
usually. Noting that the cross field fluxes are evaluated at fix
particle (as opposed to guiding center) position, these global
orderings lead to an enhancement of the gyrophase dependent
$-\vec{\rho} \cdot \nabla f$ term. This leads to that a finite angular
momentum appears already in the order we solve for $\delta f$, even
though we assume subsonic ion flows. 

From the steady-state, species-summed, flux-surface averaged momentum
conservation equation, it follows that the radial flux of toroidal
angular momentum is balanced by radial currents and/or momentum
sources.  The global simulations are not intrinsically-ambipolar, thus
if ambipolarity is not enforced radial currents arise. To be
consistent with ambipolarity these currents should be balanced by
other non-intrinsically-ambipolar processes which are outside the
physics modeled by the code (e.g. ripple losses). However, as shown in
\nouncite{buller2017}, the ambipolarity of the neoclassical radial
particle flux can be enforced. Since requiring ambipolarity $\sum_a
Z_a e\, \Gamma_a=0$ only yields one additional $\psi$-dependent
constraint (as opposed to one constraint per species), the
distribution of the momentum source among the various species
represents some degree of freedom, in addition to the freedom of
specifying the poloidal and velocity dependences of all the sources.

In the simulations presented in this paper, the momentum source is
uniform in the poloidal angle, and it is distributed to the species in
proportion to their mass; specifically, the momentum source in the
equation for species $a$, $S_{ma}$ is given by
\begin{equation}
S_{ma} = \hat{m}_a V_a(\vec{v}) S_m(\psi)
\end{equation}
where $V_a(\vec{v})=\xi x(x^2-7/2) \e^{-x^2}$, with $x=v/v_{Ta}$ and
$\xi=v_\|/v$.

\begin{figure}
  \includegraphics{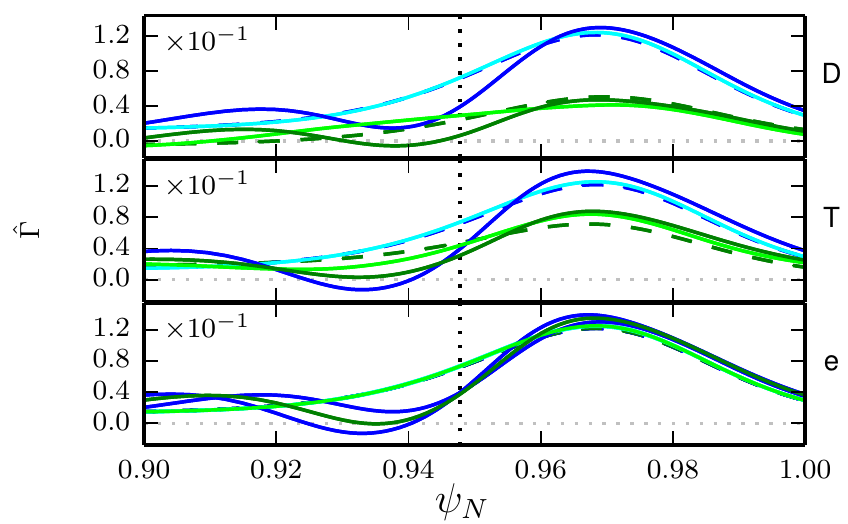}
  \caption{\label{fig:mixGamma_source2} Normalized particle flux
    $\hat{\Gamma}$ for the baseline profile with pure D and T plasma
    (blue) and a 1:1 D-T mixture (green). Solid (dashed) curves are
    from global (local) simulations. Lighter shades corresponds to
    simulations with no momentum source to electrons.}
\end{figure}

As a result of allowing for momentum source to the electrons (albeit
small in mass ratio, so that the neoclassical parallel currents are
not strongly modified), the electron particle fluxes in
\autoref{fig:baseGamma} become non-local, and are similar to the ion
fluxes in the non-intrinsically-ambipolar simulations of
Ref.~\citenum{buller2017}. On the other hand, if the momentum sources
of electrons are set to zero, the electron particle fluxes remain much
closer to the local result (shown by the bright lines in
\autoref{fig:mixGamma_source2}), as expected when collisional coupling
to the ions is weak. As a consequence, the particle fluxes in pure D
or T plasmas (light blue) are forced to be local to satisfy
ambipolarity, and become essentially isotope independent. However, in
mixture plasmas the ion components can deviate from the local result,
as only the total ion flux should be the same as the (very nearly
local) electron flux.

\bibliography{plasma-bib.bib} 
\end{document}

%% file: n_cartoon.pdf_t
\begin{picture}(0,0)%
\includegraphics{n_cartoon.pdf}%
\end{picture}%
\setlength{\unitlength}{3947sp}%
\begingroup\makeatletter\ifx\SetFigFont\undefined%
\gdef\SetFigFont#1#2#3#4#5{%
  \reset@font\fontsize{#1}{#2pt}%
  \fontfamily{#3}\fontseries{#4}\fontshape{#5}%
  \selectfont}%
\fi\endgroup%
\begin{picture}(3821,2845)(67,-1784)
\put(676,539){\makebox(0,0)[rb]{\smash{{\SetFigFont{12}{14.4}{\familydefault}{\mddefault}{\updefault}{\color[rgb]{0,0,0}$0.75$}%
}}}}
\put(676,-136){\makebox(0,0)[rb]{\smash{{\SetFigFont{12}{14.4}{\familydefault}{\mddefault}{\updefault}{\color[rgb]{0,0,0}$0.50$}%
}}}}
\put(676,-736){\makebox(0,0)[rb]{\smash{{\SetFigFont{12}{14.4}{\familydefault}{\mddefault}{\updefault}{\color[rgb]{0,0,0}$0.25$}%
}}}}
\put(676,-1336){\makebox(0,0)[rb]{\smash{{\SetFigFont{12}{14.4}{\familydefault}{\mddefault}{\updefault}{\color[rgb]{0,0,0}$0.00$}%
}}}}
\put(2176,-1711){\makebox(0,0)[lb]{\smash{{\SetFigFont{12}{14.4}{\familydefault}{\mddefault}{\updefault}{\color[rgb]{0,0,0}$\psi_N$}%
}}}}
\put(676,-1561){\makebox(0,0)[b]{\smash{{\SetFigFont{12}{14.4}{\familydefault}{\mddefault}{\updefault}{\color[rgb]{0,0,0}$0.84$}%
}}}}
\put(1426,-1561){\makebox(0,0)[b]{\smash{{\SetFigFont{12}{14.4}{\familydefault}{\mddefault}{\updefault}{\color[rgb]{0,0,0}$0.90$}%
}}}}
\put(2101,-1561){\makebox(0,0)[b]{\smash{{\SetFigFont{12}{14.4}{\familydefault}{\mddefault}{\updefault}{\color[rgb]{0,0,0}$0.96$}%
}}}}
\put(2776,-1561){\makebox(0,0)[b]{\smash{{\SetFigFont{12}{14.4}{\familydefault}{\mddefault}{\updefault}{\color[rgb]{0,0,0}$1.02$}%
}}}}
\put(3601,-1561){\makebox(0,0)[b]{\smash{{\SetFigFont{12}{14.4}{\familydefault}{\mddefault}{\updefault}{\color[rgb]{0,0,0}$1.08$}%
}}}}
\put(2251,914){\makebox(0,0)[b]{\smash{{\SetFigFont{12}{14.4}{\familydefault}{\mddefault}{\updefault}{\color[rgb]{0,0,0}$w$}%
}}}}
\put(226,-361){\rotatebox{90.0}{\makebox(0,0)[b]{\smash{{\SetFigFont{12}{14.4}{\familydefault}{\mddefault}{\updefault}{\color[rgb]{0,0,0}$n_i/(\mathrm{10^{20}m^{-3}})$}%
}}}}}
\put(2250,-462){\makebox(0,0)[rb]{\smash{{\SetFigFont{12}{14.4}{\familydefault}{\mddefault}{\updefault}{\color[rgb]{0,0,0}$n_0$}%
}}}}
\put(2476,-1186){\makebox(0,0)[rb]{\smash{{\SetFigFont{12}{14.4}{\familydefault}{\mddefault}{\updefault}{\color[rgb]{1,0,0}$n_\text{LCFS}$}%
}}}}
\put(2026,539){\makebox(0,0)[lb]{\smash{{\SetFigFont{12}{14.4}{\familydefault}{\mddefault}{\updefault}{\color[rgb]{1,0,0}$n_\text{Ped}$}%
}}}}
\put(2701,-286){\makebox(0,0)[lb]{\smash{{\SetFigFont{12}{14.4}{\familydefault}{\mddefault}{\updefault}{\color[rgb]{1,0,0}$\tfrac{\Delta n}{\Delta \psi_N}$}%
}}}}
\end{picture}%